\documentclass[12pt,preprint]{aastex}

\def\be{\begin{equation}}
\def\ee{\end{equation}}
\def\bea{\begin{eqnarray}}
\def\eea{\end{eqnarray}}

\usepackage{graphicx}

\begin{document}
\title{The magnetization degree of the outflow powering the highly-polarized reverse shock emission of GRB 120308A}
\author{Shuai Zhang\altaffilmark{1,2}, Zhi-Ping Jin\altaffilmark{1}, and Da-Ming Wei\altaffilmark{1}}
\affil{$^1$ Key Laboratory of Dark Matter and Space Astronomy, Purple Mountain Observatory, Chinese Academy of Sciences,
Nanjing, 210008, China.}
\affil{$^2$ University of Chinese Academy of Sciences,
Yuquan Road 19, Beijing, 100049, China.}
\email{jin@pmo.ac.cn,dmwei@pmo.ac.cn}

\begin{abstract}
GRB 120308A, a long duration $\gamma-$ray burst detected by {\it Swift}, was distinguished by a highly-polarized early optical afterglow emission that strongly suggests an ordered magnetic field component in the emitting region. In this work we model the optical and X-ray emission in the reverse and forward shock scenario and show that the strength of the magnetic field in reverse shock region is $\sim 10$ times stronger than that in the forward shock region. Consequently the outflow powering the highly-polarized reverse shock optical emission was mildly-magnetized at a degree $\sigma \sim$ a few percent. Considering the plausible magnetic energy dissipation in both the acceleration and prompt emission phases of the Gamma-ray Burst (GRB) outflow, the afterglow data of GRB 120308A provides us the compelling evidence that at least for some GRBs a non-ignorable fraction of the energy was released in the form of Poynting-flux, confirming the finding firstly made in the reverse-forward shock emission modeling of the optical afterglow of GRB 990123 (Fan et al. 2002; Zhang et al. 2003).
\end{abstract}

\section{Introduction}
Gamma-ray Bursts (GRBs) are brief soft $\gamma-$ray transients powered by the dying massive stars or mergers of some compact objects. In the past decades, tremendous advances in understanding such a kind of violent explosions have been
achieved \citep{Piran1999,Meszaros2006,Kumar2014}. However, the physical composition of the GRB outflows (magnetic or baryonic) is still to be better probed. In some well-studied relativistic jets for example that in active galactic nuclei, the initial ejecta are widely believed to be Poynting-flux dominated because the accretion disk is too cool to launch relativistic outflow via neutrino process (e.g., Shao et al. 2011). However, with an accretion rate high up to $\sim 1~M_{\odot}~{\rm s^{-1}}$, the accretion disk surrounding the nascent stellar-mass black hole is extremely hot and the neutrino radiation is the main cooling channel. As a result the annihilation of the neutrino and antineutrinos from the disk may be able to launch a baryonic/hot fireball \citep{Eichler1989,Piran1993,Meszaros1993}. Regardless of the nature of the central engine (either a stellar-mass black hole or a magnetized pulsar), the magnetic activity of the central engine can launch a magnetized/cold outflow \citep{Usov1992,Meszaros1997}.

Among the various methods of probing the physical composition of GRB outflows, one is to study the optical flash which is believed to be powered by the reverse shock generated in the interaction between the outflow and the circum-burst medium. The idea is the following. As long as the ordered  magnetic field component is much stronger than the random one generated by the shocks, the reverse-shock-accelerated-electrons will radiate more efficiently and then give rise to brighter optical flash. The degree of the magnetization of the reverse shock region can be inferred from the modeling of the reverse and forward emission self-consistently \citep{Fan2002,Zhang2003}. On the other hand, the synchrotron radiation of electrons in ordered magnetic field is expected to be highly-polarized and hence the optical polarimetry will be a smoking-gun signal of the magnetized outflow model \citep{Granot2003,Fan2004}. The reverse shock emission model for optical flashes \citep{Sari1999,Meszaros1999,Kobayashi2000,Shao2005} has been supported by the observations of a group of GRBs \citep[e.g.,][]{Akerlof1999,Li2003,Blake2005,Boer2006,Klotz2006,Gomboc2008,Racusin2008,Steele2009,Jin2013}. Interestingly, the modeling of almost all current optical flashes favors the mildly-magnetized reverse shock emission model \citep[e.g.,][]{Fan2002,Zhang2003,Kumar2003,Fan2005,Wei2006,Klotz2006,Gomboc2008,Racusin2008,Gao2011,Jin2013}. The absence of optical flashes in other GRBs \citep{Roming2006,Klotz2009} can be attributed to high magnetization of the outflow that can dramatically suppress the reverse shock emission \citep{Fan2004,Zhang2005,Mimica2009}. The polarimetry of the optical flashes is rather challenging. The successful polarization measurement of the a quickly-decaying optical emission of GRB 090102 got a linear polarization degree $\sim 10\%$ and has been taken as the evidence for the presence
of large-scale magnetic fields originating in the expanding fireball \citep{Steele2009}. However, a reasonable modeling of the optical and X-ray afterglow of GRB 090102 in the forward and reverse shock emission scenario was found to be not achievable \citep{Gendre2010}. The physical origin of the ``quickly-decaying optical emission" is thus less clear. Recently, \citet{Mundell2013} reported the detection of very high linear polarization degree in the early optical emission of GRB 120308A. In this work we examine whether the afterglow data is in support of the reverse shock origin of the early optical emission or not. As shown later the answer is positive, hence we estimate the magnetization degree of the outflow.

\section{Interpreting the X-ray and optical afterglow emission of GRB 120308A in the reverse-forward shock scenario}
Swift satellite triggered and located GRB 120308A, a single broad pulse of $\gamma$-rays, on 8th March 2012 at ${\rm T}_{0}=06:13:38$ UT.  The duration $T_{90}$ (15-350 keV) is $60.6 \pm 17.1$ sec and the The time-averaged spectrum from ${\rm T}_{0}-24.15$ to ${\rm T}_{0}+58.20$ sec is best fit by a simple power-law model and the power law index of the time-averaged spectrum is
$1.71 \pm 0.13$ \citep{Sakamoto2012}. The X-ray Telescope (XRT) began observing the field at $06:15:11.3$ UT, 92.6 seconds after
the BAT trigger. A bright, uncatalogued and fading X-ray source was detected \citep{Baumgartner2012}. The optical afterglow was detected by the Liverpool Telescope and other ground-based telescopes \citep{Mundell2013}. The optical emission was so bright that time-resolved polarimetry was carried out by the Liverpool Telescope with the purpose-built RINGO2 polarimeter.
At the peak time of optical emission, the linear polarization degree once reached $P=28\pm4\%$ and then declined to $P=16^{+5}_{-4}\%$ hundreds seconds later \citep{Mundell2013}. It is a very robust detection and $P=28\pm 4\%$ is the highest polarization degree of optical afterglow people have observed in all GRBs. The straightforward interpretation of the polarization properties is that the early optical emission was dominated by the reverse shock component and the outflow had large-scale uniform fields that survives long after the initial explosion, as initially identified/speculated in GRB 990123 \citep{Fan2002,Zhang2003}.

In the first 300 seconds the X-ray emission was dominated by a giant X-ray flare  with a peak $0.3-10$ keV flux $\sim 10^{-8}~{\rm erg~s^{-1}}$ which is most likely attributed to the prolonged activity of the central engine. The subsequent X-ray afterglow decayed with time as $t^{-0.704\pm0.035}$ initially and then got steeper and steeper until a jet-like break appeared at $t\sim 1.3\times 10^{4}$ s. The X-ray spectrum was $F_{\nu}\propto \nu^{-0.455^{+0.056}_{-0.081}}$ \citep{Evans2009} for $300~{\rm s}<t<10^{4}~{\rm s}$ and got significantly softened later on. In the standard fireball model, such spectral and temporal behaviors can be understood if at early times $\nu_{\rm m}< \nu_{\rm x}< \nu_{\rm c}$ and $p\sim2.1-2.2$ suppose the circum-burst medium has a constant density profile (i.e., the medium is ISM-like rather than stellar-wind like), where $\nu_{\rm c}$ ($\nu_{\rm m}$) are the cooling frequency (typical synchrotron radiation frequency) of the forward shock-accelerated electrons and $p$ is the power-law energy distribution index of the shock-accelerated electrons. In view of the spectrum change at $t\sim 1.6\times 10^{4}$ s, we expect that the cooling frequency $\nu_{\rm c}\sim 0.3~{\rm keV}$ at that time. On the other hand,
as shown in Mundell et al. (2013) the optical emission in the first $\sim 2000$ s is likely dominated by a reverse shock emission component and the forward shock emission peaked at a time $\sim 10^{3}$ s with a flux $F_{\rm opt,peak}\sim 0.3$ mJy when $\nu_{\rm m}$ crossed the observer's frequency $\nu_{\rm opt}=5\times 10^{14}$ Hz. In the ISM model, the maximal specific flux of the forward shock emission ($F_{\nu,{\rm max}}$) is a constant. Hence we have $F_{\nu,{\rm max}}=F_{\rm opt,peak}$.

\subsection{Constraining the physical parameters of the forward shock}
It is widely known that the forward shock emission is governed by the following physical parameters that can be parameterized as \citep[e.g.,][]{Piran1999,Yost2003,Fan2006}
\begin{equation}
F_{\nu,{\rm max}}=6.6~{\rm mJy}~\Big({1+z\over 2}\Big) D_{L,28.34}^{-2}
\epsilon_{B,-2}^{1/2}E_{k,53}n_0^{1/2}, \label{eq:F_nu,max}
\end{equation}
\begin{equation}
\nu_m =2.4\times 10^{16}~{\rm Hz}~E_{\rm k,53}^{1/2}\epsilon_{\rm
B,-2}^{1/2}\epsilon_{e,-1}^2 C_p^2 \Big({1+z \over 2}\Big)^{1/2}
t_{d,-3}^{-3/2},\label{eq:nu_m}
\end{equation}
\begin{equation}
\nu_c = 4.4\times 10^{16}~{\rm Hz}~E_{\rm k,
53}^{-1/2}\epsilon_{B,-2}^{-3/2}n_0^{-1}
 \Big({1+z \over 2}\Big)^{-1/2}t_{d,-3}^{-1/2}{1\over (1+Y)^2},
 \label{eq:nu_c}
 \end{equation}
where $C_p \equiv 13(p-2)/[3(p-1)]$, $\epsilon_{\rm e}$ ($\epsilon_{\rm B}$) is the fraction of shock energy given to the electrons (magnetic field),  the Compton parameter $Y\sim
(-1+\sqrt{1+4\eta \epsilon_e/\epsilon_B})/2$, $\eta \sim \min\{1,
(\nu_m/\bar{\nu}_c)^{(p-2)/2} \}$ and $\bar{\nu}_c=(1+Y)^2 \nu_c$. Here and throughout this text, the convention $Q_{\rm x}=Q/10^{\rm x}$ has been adopted.

\begin{figure}
\begin{center}
\includegraphics[width = 250pt]{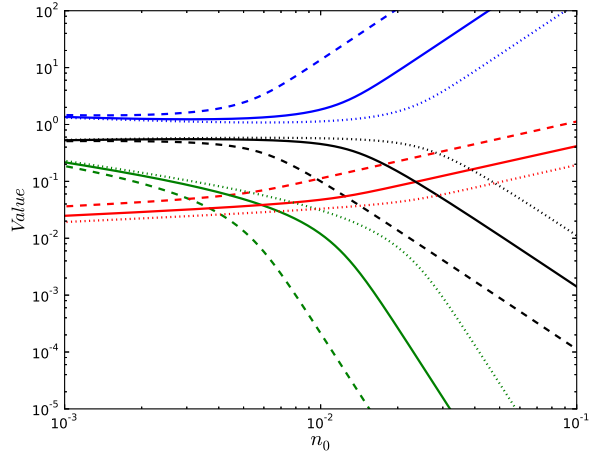}
\end{center}
	\caption{The green, red, blue and black lines represent $(\epsilon_{\rm B},~\epsilon_{\rm e}, ~E_{\rm k,53},~\eta_{\gamma})$ as a function of $n$ respectively, where $\eta_\gamma$ is the GRB efficiency. The dashed line, solid line and dotted lines correspond to $p=2.1$, $2.15$ and $2.2$, respectively.}
\end{figure}

As inferred from the optical and X-ray afterglow emission, we have $\nu_{\rm c}(t\sim 1.6\times 10^{4}~{\rm s})\approx 0.3$ keV, $\nu_{\rm m}(t\sim 10^{3}~{\rm s})\approx 5\times 10^{14}$ Hz and $F_{\nu,\rm max}\sim 0.3$ mJy, which yield
\begin{equation}
\epsilon_{\rm B,-2}^{1/2}E_{\rm k,53}n_0^{1/2}\approx a,
\end{equation}
\begin{equation}
E_{\rm k,53}^{1/2}\epsilon_{\rm B,-2}^{1/2}\epsilon_{\rm e,-1}^2 \approx b,
\end{equation}
and
\begin{equation}
E_{\rm k,53}^{-1/2}\epsilon_{\rm B,-2}^{-3/2}n_0^{-1}{(1+Y)^{-2}}\approx c.
\end{equation}
where:
\begin{equation}
a=\frac{1}{6.6}F_{\nu,{\rm max}} D_{L,28.34}^{2}\Big({1+z\over 2}\Big)^{-1}
\end{equation}

\begin{equation}
b=\frac{1}{2.4}\times 10^{-16}\nu_m C_p^{-2} \Big({1+z \over 2}\Big)^{-1/2}
t_{d,-3}^{3/2}
\end{equation}
\begin{equation}
c=\frac{1}{4.4}\times 10^{-16}\nu_c\Big({1+z \over 2}\Big)^{1/2}t_{d,-3}^{1/2}
\end{equation}

Now we have four variables while only three equations. The value of these variables can not be uniquely determined and hence we express $(E_{\rm k},~\epsilon_{\rm e},~\epsilon_{\rm B})$ into the number density of the medium $n$.

In the case of $Y\leq 1$, the term of $(1+Y)^{-2}$ can be ignored and we have
\begin{equation}
\epsilon_{B,-2}=a^{-\frac{2}{5}}c^{-\frac{4}{5}} n_{0}^{-\frac{3}{5}},
\end{equation}
\begin{equation}
\epsilon_{e,-1}=a^{-\frac{1}{5}}b^{\frac{1}{2}}c^{\frac{1}{10}} n_{0}^{\frac{1}{5}},
\end{equation}
\begin{equation}
E_{k,53}=a^{\frac{6}{5}}c^{\frac{2}{5}} n_{0}^{-\frac{1}{5}}.
\end{equation}
The dependence of all these three variables on $n$ is not sensitive.

In the case of $Y\geq 1$,  we have $(1+Y)^{-2}\approx Y^{-2}$ and then have
\begin{equation}
\epsilon_{B,-2}=(10cd)^{-\frac{8}{5p-9}}a^{-\frac{2(p-1)}{5p-9}}b^{-\frac{4(p-1)}{5p-9}} n_{0}^{-\frac{3p+1}{5p-9}},
\end{equation}
\begin{equation}
\epsilon_{e,-1}=(10cd)^{\frac{1}{5p-9}}a^{-\frac{p-2}{5p-9}}b^{\frac{3p-5}{5p-9}} n_{0}^{\frac{p-1}{5p-9}},
\end{equation}
\begin{equation}
E_{k,53}=(10cd)^{\frac{4}{5p-9}}a^{\frac{6p-10}{5p-9}}b^{-\frac{2(p-1)}{5p-9}} n_{0}^{-\frac{p-5}{5p-9}},
\end{equation}
where $d= \big (\frac{2.4}{4.4} C_{p}^{2}(\frac{1+z}{2})t_{d,-3}^{-1} \big )^{\frac{p-2}{2}} $ and $ a$, $ b $, $c $ are constant.
Comparing with the case of $Y\leq 1$, the dependence of $\epsilon_{\rm B}$ and $ E_{k} $ is rather sensitive.

For $z=2.2$ and $p=2.15$ ,we have $a=0.20$, $b=0.33 $, and $c=28.37 $ and then get
\begin{equation}
\epsilon_{B,-2}=8.48\times 10^{-9} n_{0}^{-4.26},
\end{equation}
\begin{equation}
\epsilon_{e,-1}=8.82 n_{0}^{0.66},
\end{equation}
\begin{equation}
E_{k,53}=2166.2 n_{0}^{1.63}.
\end{equation}

We have solved equations ($4-6$) numerically. As shown in Fig.1, $E_{\rm k}$ grows quickly while $\epsilon_{\rm B}$ drops sharply for $n>0.01~{\rm cm^{-3}}$ in the case of $p=2.15$. We also calculated the GRB efficiency by applying a K-correction with a reasonable factor $ k=3 $. Considering that for typical GRB the efficiency is $\eta_\gamma \geq 10\%$ and  $\epsilon_{\rm B}\leq \epsilon_{\rm e}$, we find that the value of $ n_0 $ is around 0.01 for $p=2.15$. 

\subsection{The magnetized reverse shock emission}
As already mentioned, the early optical emission is likely due to the strong reverse shock emission and the crossing time of the reverse shock is just the peak time of the optical emission (i.e., $t_{\times} \sim 300$ s). Since $t_{\times}\gg T_{90}$ (i.e., the reverse shock crossed the outflow at a time much later than the end of the prompt emission), so the fireball is thin. On the other hand, the crossing time is usually estimated as
\begin{equation}
t_{\times}\thicksim60(1+z)~{\rm s}~E_{k,54}^{1/3}n^{-1/3}_{0}\Gamma_{\rm o,2.5}^{-8/3}\sim300~{\rm s},
\end{equation}
where $\Gamma_{\rm o}$ is the initial Lorentz factor of the GRB outflow. Since $\Gamma_{\rm o}$ is very weakly dependent on $ E_{k,54} $ and $ n_{0} $, so with the above equation we know that $\Gamma_{\rm o}\sim 300$, i.e., the initial GRB outflow is ultra-relativistic.

The forward-reverse shock emission has been extensively investigated. As firstly found in GRB 990123, the physical parameters of the reverse shock can be dramatically different from that of the forward shock \citep{Fan2002,Zhang2003}. For the reverse shock emission, we usually have $\nu_{\rm m}^{\rm r}<\nu_{\rm opt}<\nu_{\rm c}^{\rm r}$ and the ratio between the reverse shock optical emission and the forward shock peak optical emission is estimated by (see eq.(16) of Jin \& Fan (2007); similar expression can be found in Zhang et al. (2003))
\begin{equation}
\frac{F^{r}_{\nu_{\rm opt}}(t_{\times})}{F_{\nu_{\rm opt}}(t_{p})} =0.08{\cal R}^{p-1}_{\rm e}{\cal R}^{(p+1)/2}_{\rm B}({\gamma_{34,\times}-1 \over 0.25})^{p-1}\Big(\frac{t_{p}}{t_{\times}}\Big)^{3(p-1)/4},
\label{eq:ratio}
\end{equation}
where $\gamma_{34,\times}\approx 1.25$ is the strength of the reverse shock emission at the crossing time (i.e., it is assumed that at that time the Lorentz factor of the decelerating outflow is half of the initial, as found in the numerical calculations), and ${\cal R}_{\rm e}=\epsilon_{\rm e}^{\rm r}/\epsilon_{\rm e}$ and ${\cal R}_{\rm B}=\sqrt{\epsilon_{\rm B}^{\rm r}/\epsilon_{\rm B}}$ ($\epsilon_{\rm e}^{\rm r}$ and $\epsilon_{\rm B}^{\rm r}$ are the fractions of reverse-shock-energy given to the electrons and magnetic field, respectively).

On the one hand, the reverse shock emission peaked at $\sim t_{\times}\sim300~{\rm s}$ and the peak flux is $F^{r}_{\nu_{\rm opt}}(t_{\times})\sim2~{\rm mJy}$. On the other hand, the forward shock optical emission likely peaked at $t\sim 10^{3}~{\rm s}$ with a flux $\sim 0.3$ mJy. With eq.(\ref{eq:ratio}) we have
\[
{\cal R}_{\rm B}\sim 10~{\cal R}^{2(1-p)/(p+1)}_{\rm e}({\gamma_{34,\times} -1 \over 0.25})^{2(1-p)/(p+1)}.
\]
In this work we assume that ${\cal R}_{\rm e}=1$ and thus ${\cal R}_{\rm B}\approx 10$ for $\gamma_{34,\times}\approx 1.25$, i.e., the reverse shock region contains magnetic field $\sim 100$ times stronger than that in the forward shock region. One reason
for this assumption is that the initially outflow is orderly magnetized or alternatively the magnetic field generated in the
internal-shock phase may have not been dissipated effectively
in a short time and would play a dominant role in the reverse shock
region. Since an ordered magnetic field is highly needed to  reproduce the rather high linear polarization detected in the reverse shock emission, we conclude that the initial outflow was magnetized. Similar conclusion was drawn in Mundell et al. (2013). However, they found a ${\cal R}_{\rm B}^{2}\sim 500$ since the smaller $\gamma_{\rm 34,\times}\approx 1.08$ was adopted in the estimate, based on \citet{Harrison2013}'s numerical calculation with the outflow spreading effect.

Let us estimate the magnetization degree of the outflow ($\sigma$), supposing the magnetic filed in the reverse shock region is dominated by the ordered component. Then at $t_{\times}$ we have
\begin{equation}
{\cal R}_{\rm B}=\sqrt{p'_{\rm r,B}/\epsilon_{\rm B}e'_{\rm f}},
\end{equation}
where $p'_{\rm r,B}$ is the comoving magnetic pressure in the reverse shock region while $e'_{\rm f}$ is the comoving thermal energy density in the forward shock region. Since $p'_{\rm r,B}/e'_{\rm f}=(p'_{\rm r,B}/p'_{\rm r,th})[(\hat{\Gamma}-1)e'_{\rm r,th}/e'_{\rm f}]$ (where $\hat{\Gamma}$ is the adiabatic index), with $p'_{\rm B}/p'_{\rm r,th}=[{\sigma}/2(\hat{\Gamma}-1)](u_{\rm us}/u_{\rm ds})(e'_{\rm r,th}/n'_{\rm r}m_{\rm p}c^{2})^{-1}$ (see eq.(12) of Fan et al. (2004b)),
we have
\begin{equation}
\sigma \approx 2{\cal R}_{\rm B}^{2}\epsilon_{\rm B}(e'_{\rm f}/n'_{\rm r}m_{\rm p}c^{3})(u_{\rm ds}/u_{\rm us}),
\end{equation}
where $n'_{\rm r}$ is the comoving number density of the reverse shock region, $u_{\rm us}$ is the velocity of the un-shocked GRB outflow relative to the surface of the reverse shock (see Fig.1(a) of Fan et al. (2004b) to see the result, note that the $\gamma_{12}$  used in Fan et al. (2004b) is just the current $\gamma_{\rm 34,\times}$), and $u_{\rm ds}$ is the velocity of the shocked GRB outflow relative to the surface of the reverse shock, which is calculated through the Lorentz transformation, i.e.,
\begin{equation}
u_{\rm us}=\gamma_{\rm ds}\gamma_{34,\times}(\beta_{\rm ds}+\beta_{34,\times}),
\end{equation}
where $\beta$ is the velocity in the unit of the speed of light $c$ and $u=\gamma \beta$. Now we have
\begin{equation}
\sigma \approx 2{\cal R}_{\rm B}^{2}\epsilon_{\rm B}{\beta_{\rm ds}\over \gamma_{34,\times}(\beta_{\rm ds}+\beta_{34,\times})}{e'_{\rm f}\over n'_{\rm r}m_{\rm p}c^{3}}.
\end{equation}
On the other hand, since $e'_{\rm f}\approx e'_{\rm r}\approx (\gamma_{34,\times}-1)n'_{\rm r}m_{\rm p}c^{2}$ unless the reverse shock region is magnetic energy dominated, we then have
\begin{equation}
\sigma \approx {\cal R}_{\rm B}^{2}\epsilon_{\rm B}{\gamma_{34,\times}-1 \over \gamma_{34,\times}}{2\beta_{\rm ds}\over (\beta_{\rm ds}+\beta_{34,\times})}.
\end{equation}
If $\beta_{\rm ds} \approx \beta_{34,\times}$, the above equation reduced to the form found in \citet{Harrison2013}, i.e., $\sigma \approx {\cal R}_{\rm B}^{2}\epsilon_{\rm B}{(\gamma_{34,\times}-1)/\gamma_{34,\times}}$. For $\gamma_{34,\times}\approx 1.25$, the fiducial value adopted in this work, we have
\[
\sigma \approx {{\cal R}_{\rm B}^{2}\epsilon_{\rm B}\over 12.5} \sim 0.01~({\cal R}_{\rm B}/10)^{2}(\epsilon_{\rm B}/0.002),
\]
where $\epsilon_{\rm B}$ is normalized to 0.002, the value obtained in our numerical fit (see section \ref{sec:Num}). 
If the outflow shell spreading effect is significant and $\gamma_{34,\times}\approx 1.08$, the magnetization is expected to be $\sigma\sim 0.03~({\cal R}_{\rm B}^{2}/500)(\epsilon_{\rm B}/0.002)$.

\subsection{Numerical fit to the data}\label{sec:Num}
\begin{figure}
\begin{center}
\includegraphics[width = 250pt]{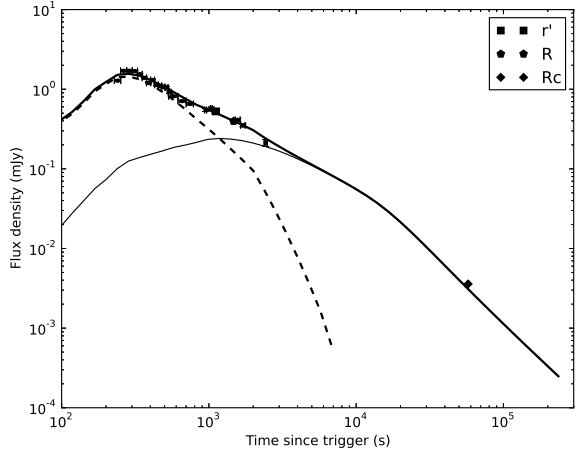}
\end{center}
	\caption{Fit of the optical data. The other data points are from Mundell et al. (2013), \citet{Virgili2012}, \citet{Elenin2012} and \citet{Bikmaev2012}. The dashed line represents reverse shock emission light curve and the dotted line represents forward shock emission light curve. The solid line is the sum of both reverse and forward shock emission.}
\end{figure}

The code used here to fit the X-ray and optical light curves has been developed by Yan
et al. (2007), in which both the reverse and the forward shock emission
have been taken into account. As already mentioned, we assume that $\epsilon_{\rm e}$ and the electron spectral
index $p$ are essentially the same for the forward shock and reverse shock, but
we allow different $\epsilon_{\rm B}$-values in these two regions.

\begin{figure}
\begin{center}
\includegraphics[width = 250pt]{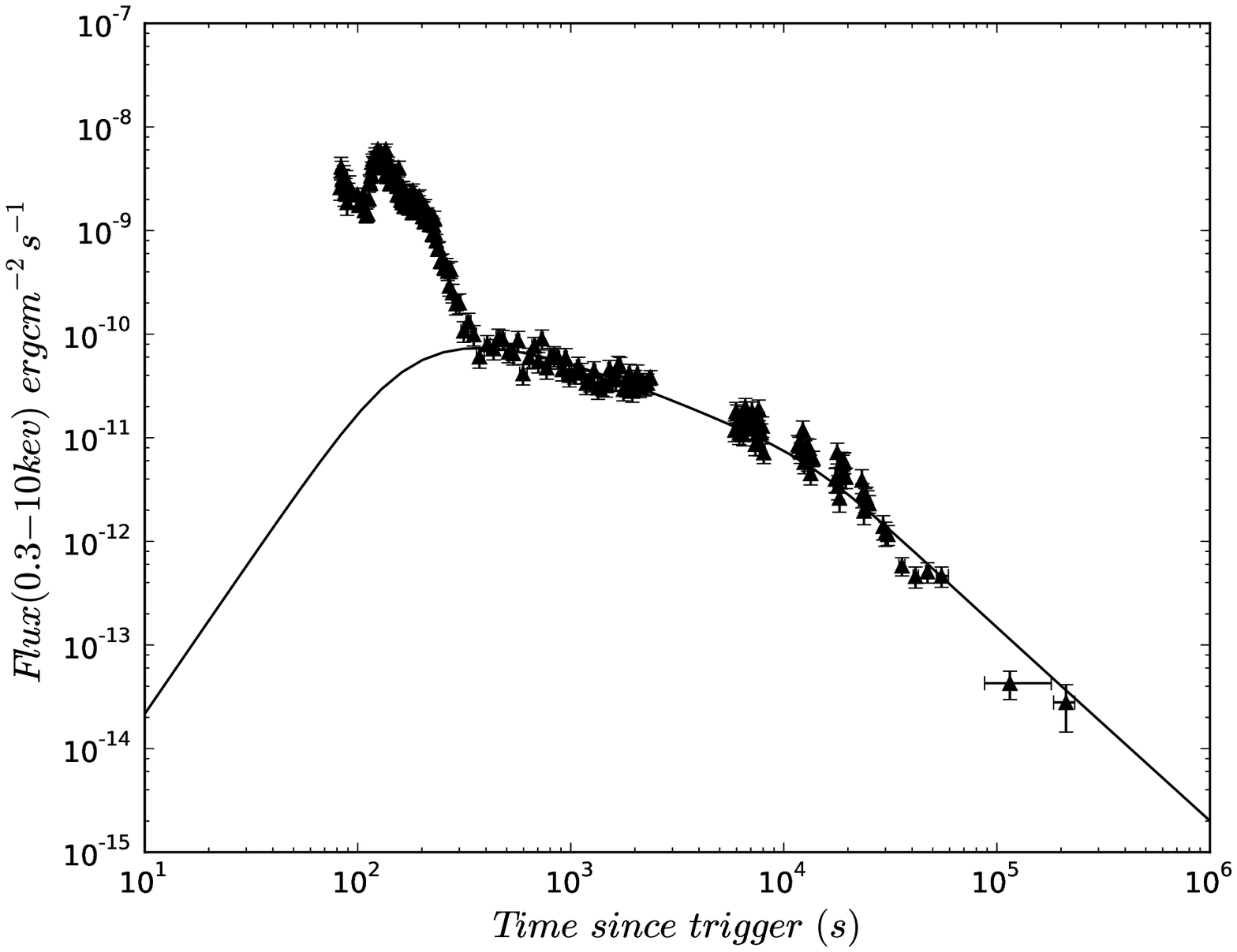}
\end{center}
	\caption{Numerical fit of X-ray data adopted from http://www.swift.ac.uk/xrt$_{-}$curves/00517234/ \citep{Evans2009}.
 }
\end{figure}

The numerical results are presented in Fig.2 and Fig.3, and the fitting parameters are $(E_{\rm k,53},~n_{0},~\epsilon_{\rm e},~\epsilon_{\rm B},~{\cal R}_{\rm B},~p,~\theta_{\rm j},~\Gamma_{0})\sim(5,~0.01,~0.05,~0.002,~7,~2.15,~0.015,~300)$, where $\theta_{\rm j}$ is the half-opening angle of the GRB outflow to account for the jet break presented in both X-ray and optical data. These parameters are well consistent with that found in our analytical estimate (see Section 2.1 and 2.2).

\section{Discussion}

The very early optical afterglow emission, in particular the bright optical flash expected in the reverse shock emission model, is very valuable to constrain the nature of the GRB outflow. This is because at such early times, the outflow likely still carries some information on the magnetization of the initial outflow. If the outflow is just weakly magnetized, there are two interestingly observational signatures: (i) the reverse shock optical emission can be significantly brightened and then outshine the forward shock optical emission (Fan et al. 2002; Zhang et al. 2003); (ii) the reverse shock optical emission will be significantly polarized and a moderate/high linear polarization is expected (Granot \& K\"{o}nigl 2003; Fan et al. 2004a). Both signals have been detected in GRB 120308A, which thus provide compelling evidence for the large scale ordered magnetic field in the initial GRB outflow (see also Mundell et al. 2013). To set a tighter constraint on the magnetization of the outflow, in this work we have modeled both the X-ray and optical emission. Due to the lack of radio detection/spectrum and then the absence of a reasonable estimate of synchrotron-self-absorption frequency of the forward shock, the shock parameters $(E_{\rm k},~n,~\epsilon_{\rm e},~\epsilon_{\rm B})$ can not be uniquely determined (see Section 2.1 for the details). Even so, if we assume a typical GRB efficiency that is expected to be not smaller than $\sim 10\%$ (note that the isotropic-equivalent $\gamma-$ray radiation energy of GRB 120308A is $\sim 6\times 10^{52}$ erg), then we have $\epsilon_{\rm B}\geq 0.002$ (see the numerical fit result). The magnetization degree of the outflow in the reverse shock region is thus $\sigma \sim $a few percent, depending on whether the outflow shell spreading effect is important or not. Considering the plausible magnetic energy dissipation in both the acceleration and prompt emission phases of the GRB outflow, we conclude that the afterglow data of GRB 120308A provides us the compelling evidence that at least for some GRBs a non-ignorable fraction of the energy was released in the form of Poynting-flux.

Finally we would like to point out that in addition to the measurement of the synchrotron-self-absorption frequency in radio bands, the degeneracy between the shock parameters $(E_{\rm k},~n,~\epsilon_{\rm e},~\epsilon_{\rm B})$ can also be broken by the observation of the synchrotron-self-Compton  GeV-TeV emission together with the optical and X-ray data because the synchrotron-self-Compton parameter $Y$ is also related to these shock parameters, too. In view of these possibilities, we urge the multi-wavelength afterglow (radio, optical, X-ray and hard $\gamma-$ray) observations of the GRBs with early optical polarimetry information.

\section*{Acknowledgments}
We thank Dr. Y. Z. Fan for stimulating discussion. This work was supported in part by 973 Programme of China under grant 2014CB845800, National Natural Science Foundation of China under grants 11273063, 11303098 and 11361140349, and the Chinese Academy of Sciences via the Strategic Priority Research Program (Grant No. XDB09000000).

\clearpage
\end{document}